\documentclass[preprint,10pt,twocolumn]{aastex63}
\usepackage[utf8]{inputenc}
\usepackage{amsmath}
\numberwithin{equation}{section}
\usepackage{amssymb}
\usepackage{mathtools}
\usepackage{enumerate}
\usepackage{enumitem}
\usepackage{array,multirow,graphicx}
\usepackage{xcolor,colortbl}
\usepackage[caption=false]{subfig}
\usepackage{float}
\usepackage{tabularx}
\usepackage{cancel}
\usepackage{ulem}
\usepackage{aas_macros}
\usepackage{hyperref}
\definecolor{Maroon}{cmyk}{0, 0.87, 0.68, 0.32}
\hypersetup{colorlinks=true, breaklinks=true, linkcolor = Maroon, citecolor=blue, urlcolor=red}

\newcommand{\code}{\texttt}
\newcommand{\dm}{\text{DM}}

\newcommand{\mmsun}{{\rm M}_\odot}

\newcommand{\mmhalo}{M_{\rm halo}}

\newcommand{\mrvir}{r_{\rm 200}}
\newcommand{\rvir}{$\mrvir$}

\newcommand{\mdmunits}{{\rm pc \, cm^{-3}}} 
\newcommand{\dmunits}{$\mdmunits$}
\newcommand{\mdmdpulsar}{{\rm \Delta DM}_{\rm pulsar}}
\newcommand{\dmdpulsar}{$\mdmdpulsar$}
\newcommand{\mdmpulsar}{{\rm DM}_{\rm pulsar}}
\newcommand{\dmpulsar}{$\mdmpulsar$}
\newcommand{\mdmcosmic}{{\rm DM}_{\rm cosmic}}
\newcommand{\dmcosmic}{$\mdmcosmic$}
\newcommand{\mdmacosmic}{\langle {\rm DM}_{\rm cosmic} \rangle}
\newcommand{\dmacosmic}{$\mdmacosmic$}
\newcommand{\mdmfrb}{{\rm DM}_{\rm FRB}}
\newcommand{\dmfrb}{$\mdmfrb$}
\newcommand{\mdmdfrb}{{\rm \Delta DM}_{\rm FRB}}
\newcommand{\dmdfrb}{$\mdmdfrb$}
\newcommand{\mdmmwism}{{\rm DM}_{\rm ISM}}
\newcommand{\dmmwism}{$\mdmmwism$}
\newcommand{\mdmdmwism}{{\rm DM}_{\rm ISM}^\delta}
\newcommand{\dmdmwism}{$\mdmdmwism$}
\newcommand{\mdmmwhalo}{{\rm DM}_{\rm MW,halo}}
\newcommand{\dmmwhalo}{$\mdmmwhalo$}
\newcommand{\mdmhost}{{\rm DM}_{\rm host}}
\newcommand{\dmhost}{$\mdmhost$}
\newcommand{\mdmmhost}{{\rm DM}_{\rm host}^{\rm min}}
\newcommand{\dmmhost}{$\mdmmhost$}

\def\ion#1#2{{#1}\,{\sc #2}}

\newcommand{\ovi}{\ion{O}{vi}}
\newcommand{\ovii}{\ion{O}{vii}}

\begin{document}

\title{A Data-Driven Technique Using Millisecond Transients to Measure the Milky Way Halo}

\author{E. Platts}
\affiliation{High Energy Physics, Cosmology \& Astrophysics Theory (HEPCAT) group, Department of Mathematics and Applied Mathematics, University of Cape Town, South Africa}

\author{J. Xavier Prochaska}
\affiliation{Department of Astronomy \& Astrophysics, UC Santa Cruz, USA}
\affiliation{Kavli Institute for the Physics and Mathematics of the Universe (Kavli IPMU; WPI), The University of Tokyo, Japan}

\author{Casey J. Law}
\affiliation{Department of Astronomy and Owens Valley Radio Observatory, California Institute of Technology, Pasadena, CA 91125, USA}

\date{May 2020}

\begin{abstract}
We introduce a new technique to constrain the line-of-sight integrated electron 
density of our Galactic halo \dmmwhalo\ through analysis of the
observed dispersion measure distributions of pulsars 
\dmpulsar\ and fast radio bursts \dmfrb.
We model these distributions, correcting for the Galactic
interstellar medium, with kernel density estimation---well-suited to the small data regime---to find lower/upper bounds
to the corrected \dmpulsar/\dmfrb\ distributions:
$\max\left[\dm_{\rm pulsar}\right] \approx 7\pm2 \, \text{(stat)} \pm 9 \, \text{(sys)} \, \mdmunits$ 
and
$\min \left[\dm_{\rm FRB}\right] \approx 63^{+27}_{-21}  \, \text{(stat)} \pm 9 \, \text{(sys)} \, \mdmunits$.
Using bootstrap resampling to estimate uncertainties,
we set conservative limits on the Galactic halo dispersion measure 
$-2 < \mdmmwhalo < 123 \, \mdmunits$  (95\% c.l.).
The upper limit is especially conservative because it may include a 
non-negligible contribution from the FRB host galaxies and
a non-zero contribution from the cosmic web.
It strongly disfavors models where the Galaxy has retained
the majority of its baryons with a density profile tracking
the presumed dark matter density profile.
Last, we perform Monte Carlo simulations of larger FRB samples
to validate our technique and assess the sensitivity of ongoing
and future surveys.  We recover bounds of several tens \dmunits\ 
which may be sufficient to test whether the Galaxy has 
retained a majority of its baryonic mass. We estimate that a sample of several thousand FRBs will significantly tighten constraints on \dmmwhalo\ and offer a valuable complement to other analyses.

\end{abstract}

\section{Introduction}
 
In the early universe the majority of baryons resided in a cool, diffuse plasma, which is predicted to have collapsed into sheetlike and filamentary structures that make up the intergalactic medium (IGM). Around the time of structure formation, dark matter collapses into halos, pulling baryons with it. As the gas falls inwards, it is shock-heated to form a hot, diffuse plasma, known as halo gas or the circumgalactic medium (CGM). Approximately $10\%$ of the gas cools and falls into the center of the halo to form stars and the interstellar medium 
\citep[ISM; e.g.][]{White:1977jf}. 

Comparing the baryonic mass fraction detected for galaxies ($M_b/M_{\text{halo}}$) to the cosmic mean ($\Omega_b/\Omega_m$), however, reveals a baryonic deficit \citep[e.g.][]{Dai_2010}. The missing baryons may have been ejected back into the IGM before forming stars 
or perhaps have yet to be detected 
\citep[e.g.][]{2011ApJ...740...91P,10.1111/j.1365-2966.2011.20047.x}. In the latter scenario, the CGM presents itself as a possible refuge.

This issue holds for the CGM of our Galaxy.  While it is evident that its stars and ISM correspond to $\lesssim 25\%$ of the baryonic mass available to a halo with mass $\mmhalo = 10^{12.2} \mmsun$
\citep[the current estimate;][]{boylin}, 
the mass and distribution of gas within our Galactic halo are not well determined even despite our close proximity. The key observables that constrain the Galactic CGM include soft X-ray emission from the plasma \citep{henley2010}, X-ray and UV absorption-lines of oxygen ions \citep{ovi,ovii}, density constraints from ram-pressure stripping of the Large Magellanic Cloud \citep[LMC;][]{salem+2015}, and dispersion measure (DM) observations from pulsars towards the LMC \citep{2006ApJ...649..235M}. These have provided valuable constraints for models of the Galactic halo, 
but still allow for large variations in the mass and 
spatial extent of the gas \citep{fang,bregman,faerman,xyz19}. 

A primary challenge to assessing the Galactic CGM is that the gas is too diffuse (especially at large radii) to be imaged directly.  Furthermore, the absorption-line measurements 
(e.g.\ \ovi\ and \ovii) require substantial ionization and/or metallicity corrections to infer the total gas. In this respect, the DM measurements towards the LMC provide the most direct probe of the ionized gas, yet it lies at only $\approx 1/4$ the virial radius \rvir\ of the Galaxy. 
Ideally, one would prefer to record
DM measurements to \rvir\ and also
across the sky to search for asymmetries in the halo gas distribution. 
Just such an opportunity is now afforded (albeit with caveats, as we will discuss) by the transients known as fast radio bursts (FRBs).

FRBs are the population of $\sim$millisecond chirps of bright
radio emission at approximately GHz frequencies discovered
serendipitously \citep{Lorimer777} and now pursued in earnest
with dedicated projects and facilities \citep{2016MNRAS.458..718C, 2018ApJS..236....8L,2018ApJ...863...48C,2019MNRAS.489..919K}.
Recorded in each FRB event is its DM value \dmfrb. The majority greatly exceed estimates for our Galactic ISM and CGM,
lending strong statistical support that FRBs have an extragalactic origin \citep{Petroff:2019tty,2019ARA&A..57..417C}.
This inference has been confirmed by a small but growing set of FRBs localized to $\approx 1''$ and then shown to reside in a distant galaxy \citep{Tendulkar:2017vuq,2019Sci...365..565B,Ravi:2019alc,Prochaskaeaay0073,2020Natur.577..190M}. As a result, the community now recognizes FRBs as a viable tool to probe ionized gas across the universe, e.g.\ to conclusively detect the so-called ``missing'' baryons of the present-day universe \citep{1998ApJ...503..518F,2018NatAs...2..836M}.

Owing to its integral nature, \dmfrb\ includes contributions from all of the electrons along the sightline:  the intergalactic medium, gas in distant Galactic halos, the ionized gas of the system hosting the FRB, and our Milky Way. Indeed, the host and Galaxy contributions (\dmhost, DM$_{\rm MW}$) are frequently considered a ``nuisance'' to proposed analyses of the cosmic web. In this manuscript, however, we view them as a highly desired signal, i.e.\ a new opportunity to constrain the Galactic CGM.

There are two primary challenges that this paper addresses: how to use pulsars and FRBs to probe the dispersion measure of Galactic halos, and how to do so with a limited data set. The first problem is addressed by
constraining the DM contribution of the MW halo to the total observed DM of pulsars and FRBs. 

For the second challenge, only $\sim 100$ FRBs have been observed to date; this necessitates techniques that are well suited to dealing with small data sets. We propose the use of standard kernel density estimation \citep[KDE;][]{Silverman86} and asymmetric, variable-bandwidth KDE \citep{chen2000,hoffmann2015unified} to find probability density functions (PDFs) of the DM distribution of pulsars and of FRBs, respectively. Other density estimation techniques are explored---namely, density estimation using field theory \citep[DEFT;][]{PhysRevE.90.011301,PhysRevE.92.032107,PhysRevLett.121.160605} and a generalized extreme value (GEV), but prove to be insufficient (see \S~\ref{sec:deft} and \S~\ref{sec:gev} for details). From the PDFs one can estimate the maximum MW halo DM given by pulsars, and the minimum MW halo and host halo DM given by FRBs.
This infers constraints on the DM of the MW CGM and part of the host CGM. 

We measure a MW halo DM of $63^{+27}_{-21} \, (\text{stat}) \pm 9 \, (\text{sys})  \, \mdmunits$, corresponding to a 1$\sigma$\ confidence detection.  The precision of this measurement is limited by the FRB sample size and we predict a robust detection of the MW halo with the incorporation of FRB detections anticipated in the coming year. The techniques presented here will make the best precision and least ambiguous measurement of the MW halo in several years with samples of $10^4$\ FRBs.

The paper is structured as follows. \S~\ref{sec:framework} outlines the core concepts of this work. \S~\ref{sec:kde} details the density estimation techniques used in the analysis. The methodology and results are presented in \S~\ref{sec:methodology}, where \S~\ref{subsec:obs} provides constraints based on observed data and \S~\ref{subsec:sim} provides an analysis based on simulations. The results and implications are discussed in \S~\ref{sec:discussion}, and conclusions are summarized in \S~\ref{sec:conclusion}.

%%%%%%%%%%%%%%%%%%%%%%%%%%%%%%%%%%%%%%%%%%%%%%%%%%%%%%%%%%%%%%%%%%%%%%%
%%%%%%%%%%%%%%%%%%%%%%%%%%%%%%%%%%%%%%%%%%%%%%%%%%%%%%%%%%%%%%%%%%%%%%%
\section{The Framework}
\label{sec:framework}
Pulsars and FRBs are both millisecond radio transients. The former lie in the disk of the MW galaxy and the latter are extragalactic. Since the group velocity of the electromagnetic wave depends on the free electron density ($n_e$) along the path of propagation, the arrival time
of the transient signal is extended. 
This spread is described by the dispersion measure:
\begin{equation}
\dm = \int \frac{n_eds}{1+z}.
\end{equation}
DMs can therefore be used to study the distribution of baryons along the line of sight between a transient source and an observer. 

\begin{figure*}
	\includegraphics[width=\linewidth]{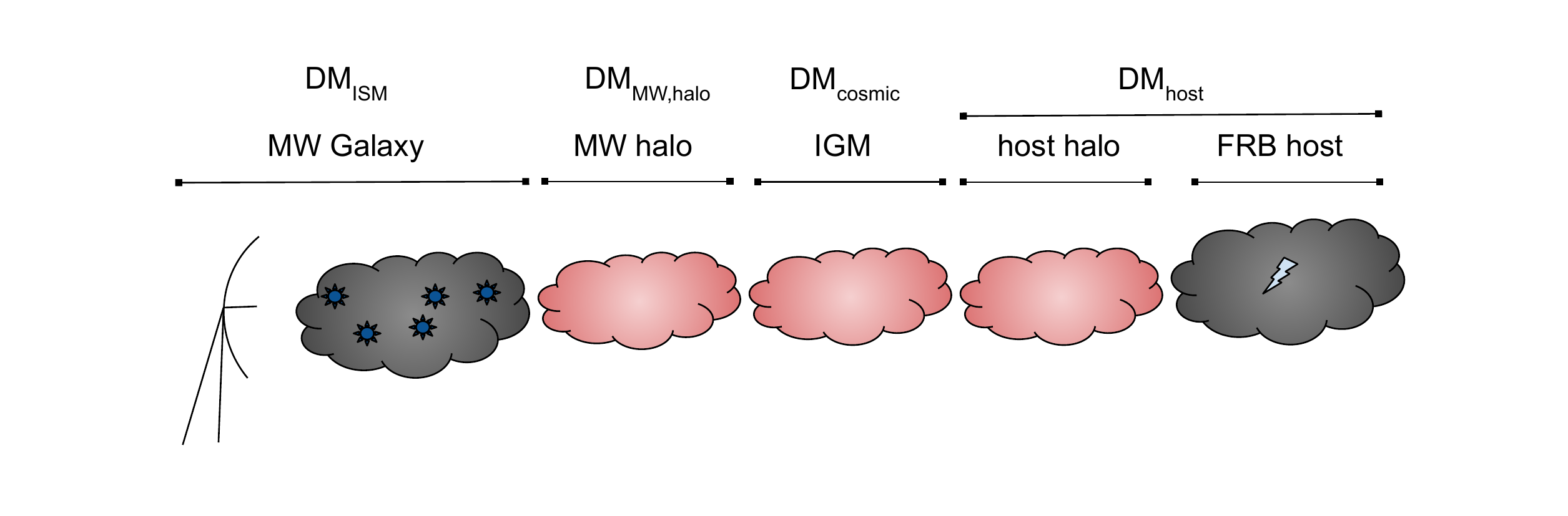}
	\caption{Schematic of the radio telescope (left-most image), the distribution of electrons (cloud shapes) that contribute to DM, and the millisecond transients (sun and lightning symbols) that are used to measure the DM. The regions shown in red have electrons, but no sources of millisecond transients. For sources distributed throughout their host galaxies and host galaxies distributed over a range of distances, the minimal Milky Way, IGM and FRB host DM contributions are zero.
	\label{fig:ne}}
\end{figure*}

Figure \ref{fig:ne} shows a schematic of how electrons are distributed relative to pulsars and FRBs. Galactic halos are assumed to be devoid of radio transients, but contain a significant column density of electrons. 
Pulsars have been detected predominantly in the Galactic disk or 
nearby globular clusters\footnote{The more distant pulsars purported to reside in
the Magellanic clouds \citep[e.g.,][]{2013MNRAS.433..138R} 
are excluded from this analysis.} 
\citep{2005AJ....129.1993M}.
Those with known distance have been used to create detailed models of the electron density distribution of the Milky Way disk \citep{Cordes:2002wz,Cordes:2003ik,2008PASA...25..184G,2017ApJ...835...29Y}. In the following we adopt both the NE2001\footnote{Available in 
Python at \href{https://github.com/FRBs/ne2001}{https://github.com/FRBs/ne2001}} and YMW16\footnote{Available in Python at
\href{https://github.com/telegraphic/pygedm}{https://github.com/telegraphic/pygedm}} algorithms.

If we assume FRBs are distributed throughout their host galaxies and throughout space, then the lowest \dmfrb\ values set a bound on the electron column density associated with the halos of the Milky Way and the typical host galaxy. This measurement is the focus of the manuscript.
Table \ref{tab:notation} provides a summary of the notation used in this paper.

\begin{table}[ht]
\footnotesize
\captionsetup{font=footnotesize}
\centering
\begin{tabularx}{\linewidth}{c|c}
Quantity & Description \\
\hline
$\dm_\text{pulsar}$ & The total DM measurement of a pulsar \\
$\dm_\text{FRB}$ & The total DM measurement of an FRB \\
\dmdmwism & DM from a fraction of the Galactic ISM \\
\dmmwism & Total sightline DM for the Galactic ISM\\
$\dm_\text{MW,halo}$ & DM of all gas in our Galactic halo \\
$\dm_\text{MW,halo}^\delta$ & DM from a fraction of gas in our Galactic halo \\
$\dm_\text{IGM}$ & DM from the IGM (gas between halos) \\
$\dm_\text{cosmic}$ & DM from all cosmic gas (IGM+halos) \\
$\mdmacosmic$ & Average DM from all cosmic gas \\
$\dm_\text{host}$ & DM from FRB host galaxy halo \\
\end{tabularx}
\caption{Notation\label{tab:notation}}
\end{table}

%%%%%%%%%%%%%%%%%%%%%%%%%%%%%%%%%%%%%%%%%%%%%%%%%%%
\subsection{Constraints from Pulsars}

We consider

\begin{equation}
\label{eq:pulsar}
    \mdmpulsar = \mdmdmwism\ + \mdmmwhalo^\delta \;\; ,
\end{equation}
with \dmdmwism\ the ISM contribution
and $\mdmmwhalo^\delta$ the halo contribution.
We then define an ISM-corrected quantity \dmdpulsar, which
subtracts the total ISM contribution 
along the pulsar sightline,
\begin{equation}
\label{eq:deltadmpsr}
    \mdmdpulsar = \mdmpulsar - \mdmmwism \;\; .
\end{equation}
Most pulsars have unknown distances yet are expected to
lie predominantly in the Galactic disk, with a scale height of $100 \, \text{pc}$ \citep{2006ApJ...643..332F}. 
Therefore, \dmmwism\ is generally larger than \dmpulsar\ and the majority of \dmdpulsar\ values will be negative.
Any positive values could be attributed to the halo,
and therefore the maximum \dmdpulsar\ yields a lower limit:

\begin{equation}
    \label{eq:delta_psr}
    \dm_\text{MW,halo} > \max\left[\mdmdpulsar\right] \;\; .
\end{equation}
Such an analysis must allow for uncertainties in 
the modeling of \dmmwism, but for high Galactic latitudes
these uncertainties
are expected to be less than $10\, \mdmunits$.

%%%%%%%%%%%%%%%%%%%%%%%%%%%%%%%%%%%%%%%%%%%%%%%%%%%%%
\subsection{Constraints from FRBs}
$\dm_\text{FRB}$ has contributions from the ISM, the MW halo, 
cosmic gas, and the FRB host galaxy,
\begin{equation}
\label{eq:frb}
    \mdmfrb = \mdmmwism +\dm_\text{MW,halo}
    + \mdmcosmic + \mdmhost \;\; .
\end{equation}
Similar to the pulsars, we define an ISM-corrected measure:
\begin{equation}
\label{eq:deltadmfrb}
    \Delta\dm_\text{FRB} = \dm_\text{FRB}-\dm_\text{ISM} \;\; .
\end{equation}
From the full distribution of \dmdfrb, we will 
examine the lowest values on the expectation that these have lower
\dmcosmic\ contributions.  For reference, an FRB at $z=0.03$
\citep[e.g.][]{2020Natur.577..190M} has an
average $\mdmacosmic \, \approx \, 25 \mdmunits$.

The lowest values of \dmdfrb\ should also reflect the lowest
combinations of \dmmwhalo\ and \dmhost.
We expect significant variations in the latter both due to the 
distribution of host galaxy masses and also from variations
in the FRB location within the galaxy.  
We express \dmmhost\ as the minimum of
this distribution which may be 10 to several tens \dmunits.

Regarding variations in \dmmwhalo, galaxy formation models tend to predict a nearly
spherical distribution of gas, especially 
beyond the inner halo (but see \citet{Yamasaki_2020} which includes a nonspherical component).  Spherically symmetric models of our Galaxy yield less than $10 \, \mdmunits$ variations in \dmmwhalo\ even though the Sun is located off-center \citep{xyz19}. In the following, we will assume a single \dmmwhalo\ unless otherwise discussed.
One recovers

\begin{equation}
\label{eq:delta_frb}
    \mdmmwhalo + \mdmmhost = \min\left[\mdmdfrb \right] \;\; ,
\end{equation}
and therefore
\begin{equation}
\mdmmwhalo < \min\left[\mdmdfrb\right] \;\; .
\end{equation}

%%%%%%%%%%%%%%%%%%%%%%%%%%%%%%%%%%%%%%%%%%%%%%%%%%%%%%%%%
\section{Kernel Density Estimation}
\label{sec:kde}
KDE is a non-parametric technique that estimates an unknown density by constructing a kernel at each data point and summing their contributions. Owing to their shapes, the distributions of \dmdpulsar\ and \dmdfrb\ are each suited to a different class of KDE. \dmdpulsar\ has smooth edges and can be adequately modelled with a Gaussian kernel and a fixed bandwidth. The sharp edge of \dmdfrb, however, necessitates a varying bandwidth and a kernel with a steep cut-off.

In \S~\ref{sec:stdkde} we outline standard KDE and in \S~\ref{sec:bvkde} we describe the modifications for asymmetric, bandwidth-varying KDE.

\subsection{Standard KDEs}
\label{sec:stdkde}
Consider an independent and identically distributed sample $\{X_i : i = 1, ..., n \}$ drawn from some unknown distribution $f(x)$. We wish to obtain an estimate $\hat{f}(x)$ of this distribution using KDE:
\begin{equation}
\hat{f}(x) = \frac{1}{n}\sum_{i=1}^n K_h\left(X_i-x\right) = \frac{1}{nh}\sum_{i=1}^n K \left(\frac{X_i-x}{h}\right) \;\;,
\end{equation}
where $K$ is the kernel and $h>0$ is the bandwidth. The kernel is the underlying distribution function and the bandwidth is a smoothing parameter. In standard KDE symmetric kernels are used, such as Gaussian, triangular, cosine, biweight, triweight, or Epanechnikov. While an Epanechnikov kernel is most optimal in terms of the mean squared error, a Gaussian kernel is the most widely used: the loss of efficiency is marginal ($\sim 5\%$) and the distribution offers convenient mathematical properties. As such, a Gaussian kernel is used in our analysis of \dmdpulsar.  Bandwidth selection is a trade-off between the bias of the KDE and its variance.  Often the bandwidth is chosen to minimize the mean integrated squared error (MISE),
\begin{equation}
    \text{MISE}(h) = \text{E}\left[\int\left(\hat{f}(x) - f(x) \right)^2 dx \right] \; \; ,
\end{equation}
which is equivalent to the expected $L_2$ risk function. $f(x)$ is unknown, however it can be approximated through various techniques (see \citet{10.2307/2291420}). One can also use rule-of-thumb bandwidth estimators, such as Silverman's \citep{Silverman86} and Scott's \citep{10.1093/biomet/66.3.605}, however these assume the underlying distribution is Gaussian. In our analysis we use \code{scikit-learn} to select the optimal bandwidth via cross-validation.

The \code{KernelDensity()} function invokes a nearest neighbors based approach: instead of using the full data set to estimate the density at each point, a number of neighboring points are selected based on the bandwidth. This improves the algorithm efficiency by ignoring distant points that have a negligible effect. KDEs are generated for a range of bandwidths, and \code{GridSearchCV()} is used to find the optimal bandwidth. Here $n$-fold cross-validation is performed. The pulsar data is divided into $n$ subsets, a KDE is generated using the data from $n-1$ subsets (training data), and the performance of the KDE is evaluated on the remaining subset (test data) by calculating the log-likelihood, $\sum \log \hat{p}(x_i)$. This process is repeated $n$ times, using a different subset as the test set each time, to give a final (averaged) log-likelihood score. In this manner, scores are calculated for a range of bandwidths. The bandwidth with the maximum log-likelihood is selected for the analysis ($h \approx 10$).

\subsection{Asymmetric KDEs}
\label{sec:bvkde}
Standard KDE performs well when the underlying distribution is unbounded and the density of data is relatively uniform. 
We will show, however, that the
\dmdfrb\ distribution has data concentrated 
towards the front of the distribution and is bounded on $[0,\infty)$. This presents two problems  that standard KDE cannot resolve. Firstly, a fixed bandwidth $h$ entails a trade-off between large and small scale structure: over-dense regions will be over-smoothed by a large $h$, and under-dense regions will be over fitted if  $h$ is too small. Secondly, symmetric kernels have significant bias at or near a boundary, known as edge or boundary effects. A fixed and symmetric kernel will allocate weight outside of the density region when smoothing the distribution. 

Various techniques have been developed that attempt to resolve this issue, eg. data reflection \citep{Schuster:1985},
boundary kernels \citep{muller:1991,muller:1993,muller:1994}, the hybrid method \citep{hall:1991}, generating pseudo-data \citep{cowling:1996}, data binning and local polynomial fitting \citep{cheng:1997}, and others. One can also invoke asymmetric kernels (such as gamma, lognormal and inverse Gaussian) and variable bandwidths. In this work we use gamma estimators developed by \cite{chen2000} and expanded upon by \cite{jeon2013} and \cite{hoffmann2015unified}.

The gamma PDF with standard gamma function $\Gamma(\cdot)$ is given by

\begin{equation}
    K_{k,\theta}(x) = \frac{x^{k-1}\exp(-\frac{x}{\theta})}{\theta^k\Gamma(k)} \; \; ,
\end{equation}
with scale parameter $k$ and shape parameter $\theta$. \cite{chen2000} take $k=\rho_h(x)$ and $\theta=h$ with random gamma variables $X_i$ to obtain

\begin{equation}
    K_{\rho_h(x),h}(X_i) = \frac{X_i^{\rho_h(x)-1}\exp(-\frac{X_i}{h})}{h^{\rho_h(x)}\Gamma(\rho_h(x))} \; \; ,
\end{equation}

with

\[
    \rho_h(x) = \begin{dcases*}
      \; \frac{x}{h} \, , & if \hspace{0.5cm} $x \geq 2h $ \; , \\
      \left(\frac{x}{2h}\right)^2 + 1 \, , & if \hspace{0.5cm} $x \in [0,2h)$ \; \; .
    \end{dcases*}
\]

The resulting gamma estimator is given by
\begin{equation}
    \hat{f}(x) = \frac{1}{n}\sum_{i=1}^n K_{\rho_h(x),h}(X_i) \; \; .
\end{equation}

The shape of gamma kernels vary naturally,  allowing for different smoothness at different points of the distribution. Further, because gamma kernels are non-negative, the gamma estimator itself is unlikely to deviate below zero. The bandwidths $h$ depend either on the point of estimation ($h(x)$; a balloon estimator), or on the sample associated with a kernel ($h(X_i)$; sample-smoothing estimator). In this analysis we consider the former.

Another challenge for standard KDEs is that regions with few samples have overestimated densities and regions with many are underestimated. Shifted KDEs minimize this bias by moving samples from higher to lower density regions. Combining this with balloon estimators \citep{hoffmann2015unified}, one has

\begin{equation}
    \hat{f}(x) = \frac{1}{n}\sum_{i=1}^n K_{\rho_h(x),h(x)}\left(X_i - h^p(x)\delta(x)\right) \; \; ,
\end{equation}

where $p$ is the order of the kernel and $\delta(x)$ is the shift. The kernel is shifted by $h^p(x)\delta(x)$, which vanishes for small bandwidths. For our analyses, we use Python code by \cite{hoffmann2015unified}\footnote{Available at \url{https://github.com/tillahoffmann/asymmetric_kde}}, where the optimal bandwidth for each kernel is chosen by minimizing the MISE.

%%%%%%%%%%%%%%%%%%%%%%%%%%%%%%%%%%%%%%%%%%%%%%%%%%%%%%%%%
\section{Methodology and Results}
\label{sec:methodology}

\subsection{Bounding the DM Distributions}
As described in \S~\ref{sec:framework},
we wish to estimate a maximum \dmdpulsar\ and a 
minimum \dmdfrb\ from the observed distributions.
We will first apply the appropriate formalism to derive a PDF
for each.  The minimum/maximum of the PDF, however, is
not a precisely posed quantity.
Here we introduce a metric tailored primarily for \dmdfrb\ as an estimator after
experimenting on simulated distributions (\S~\ref{subsec:sim} and \S~\ref{sec:moresim}):
the maximum gradient of the distributions, $\max\left[ f'(\Delta\dm)\right]$.
This approach is based on the physical prior that the \dmfrb\ distributions will have sharp cut-offs, which will hold if the variance in 
\dmmwhalo\ is much less than its average.  
It is further supported by the current set of FRB observations. The observed \dmdpulsar\ PDF, on the other hand, is more evenly distributed with smoother edges. As such, estimates for $\max\left[\Delta\dm_\text{pulsar}\right]$ given by the metric are more conservative. This effect is discussed in \S~\ref{sec:moresim}.

In \S~\ref{subsec:obs}, KDE analysis is performed on observed transient samples to place current constraints on $\dm_\text{MW,halo}$
from \dmdpulsar\ and \dmdfrb.
In \ref{subsec:sim}, the KDE (gamma) methodology is analysed by simulating \dmdfrb. Random samples of size $n=100,1000$ and 10,000 are taken and $\min\left[\Delta\dm_\text{FRB,sim}\right]$ 
compared to the known inputs.
This analysis also offers insight into the statistical
power of future samples.

\subsection{Observed Sample}
\label{subsec:obs}
To define our sample of pulsars and FRBs, we use the largest aggregation sites for each type of object. For pulsars, we downloaded the ATNF pulsar catalog \citep[version 1.61;][]{2005AJ....129.1993M}. For FRBs, we downloaded the FRBCat \citep[downloaded 25 February 2020, verified events only;][]{2016PASA...33...45P}. 

The Milky Way electron distribution is more complex at low Galactic latitudes owing to contributions from spiral arms, HII regions,
and supernova remnants. Electron density models are most complex on size scales smaller than 200 pc and within 1 kpc of the Sun \citep{Cordes:2003ik}. To minimize systematic error introduce by the model, we only consider sources more than $200/1000\approx20$\ deg from the galactic plane; we also compare the results with a second, more conservative cut to estimate systematic error.
We also remove all pulsars within 5 deg of the Magellanic clouds. For a latitude limit of $|b|>20$\,deg, the samples include $371$ pulsars and $83$ FRBs. For a latitude limit of $|b|>30$\,deg, the
samples include $215$ pulsars and $64$ FRBs. Owing to the significant decrease in FRB data for $|b|>30$\,deg, the final results presented in this paper use a Galactic cut of $|b|>20$\,deg. 

This analysis requires correcting by the total \dmmwism\ contribution estimated from the Milky Way. 
Even at high Galactic latitudes, the electron density models have systematic uncertainties on the order of tens of percent due to modeling errors \citep{2012MNRAS.427..664S}. We estimate \dmmwism\ with both the NE2001 \citep{Cordes:2002wz,Cordes:2003ik} and YMW16 \citep{2017ApJ...835...29Y} models as a way of estimating potential systematic errors.

\begin{figure}[tb]
\subfloat{
	\includegraphics[width=\linewidth]{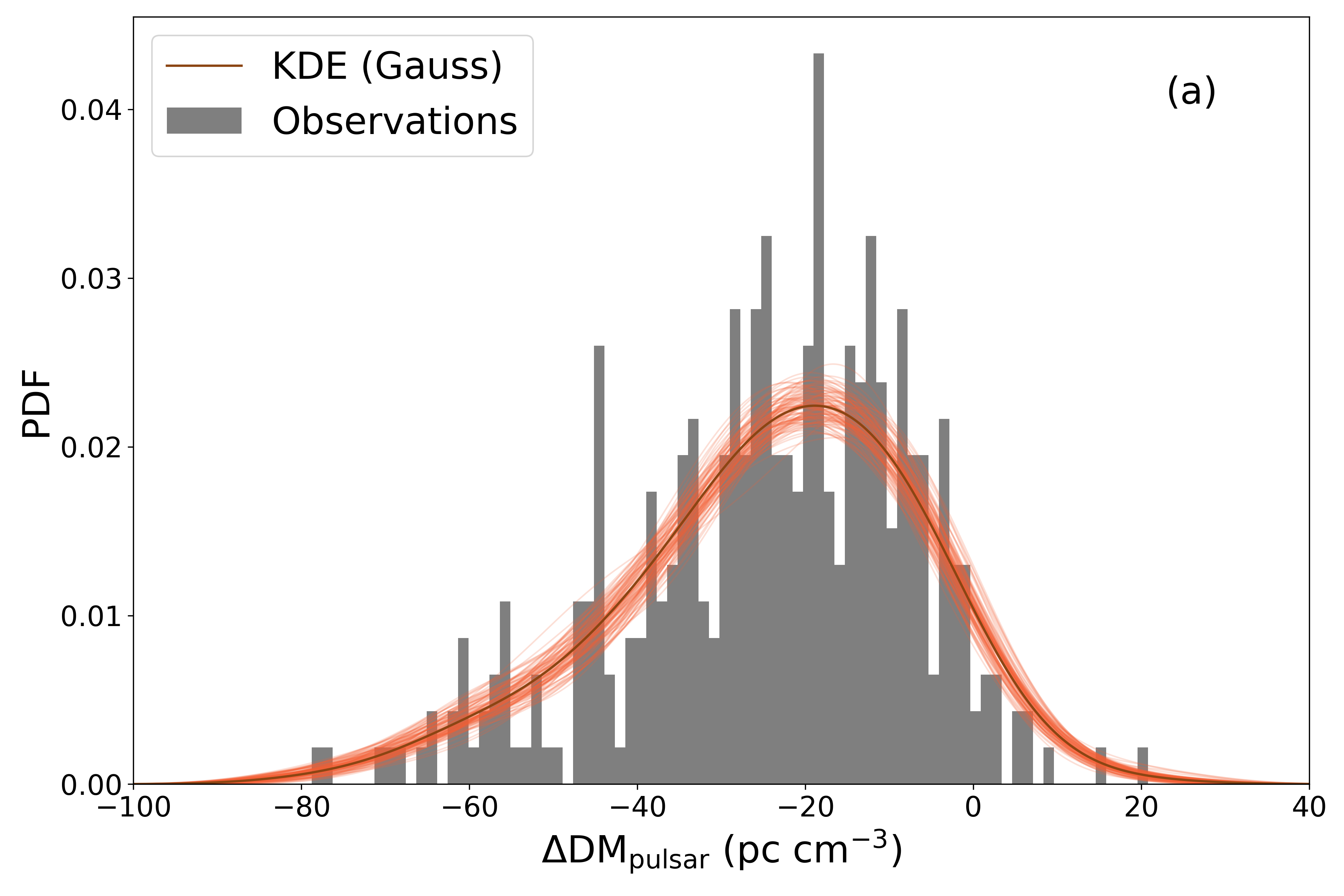}
	\label{fig:dmpsr}}

\subfloat{
	\includegraphics[width=\linewidth]{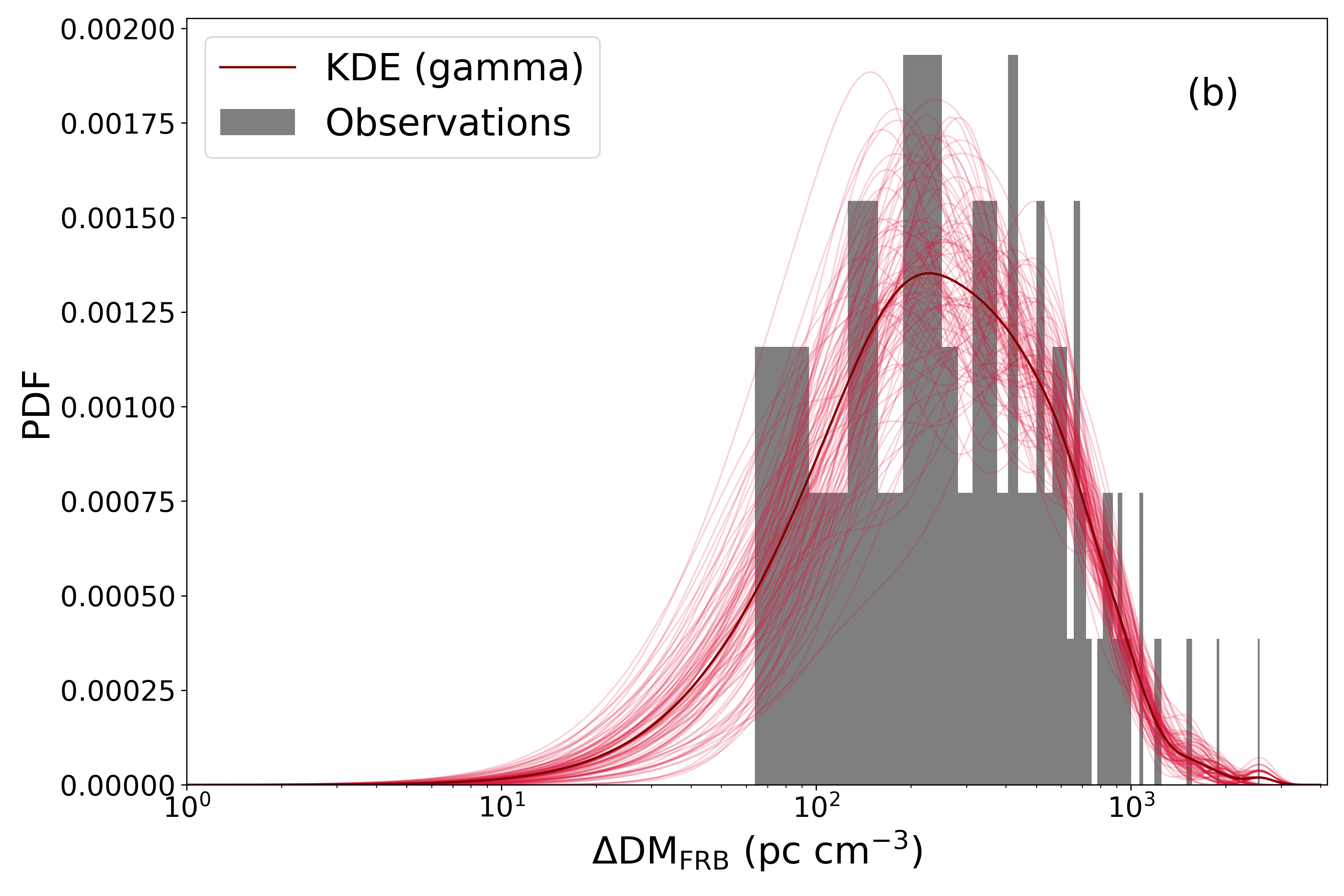}
	\label{fig:dmfrb}}

\caption{
Distributions for observed samples, restricted to $|b| > 20$\,deg and using
NE2001 for modeling \dmmwism. Overlaid on the data are PDFs derived with KDE.
(a) \dmdpulsar\ KDEs (with Gaussian kernels and a fixed bandwidth) overlaid on the observed data. The dark orange curve denotes the PDF estimated with the original data, and the lighter curves denote PDFs generated with resampled data. The bandwidth for each distribution is selected with cross-correlation and a search range between $h=8$ and $h=15$.
 (b) \dmdfrb\ KDEs (with gamma kernels and variable bandwidths) overlaid on the observed data. The thick dark red curve denotes the PDF generated with the original data and the lighter curves denote PDFs generated with the resampled data.
}
\end{figure}

We then generated distributions of 
\dmdpulsar\ and \dmdfrb, as given by 
Equations \ref{eq:deltadmpsr} and
\ref{eq:deltadmfrb}.
These are shown in Figure~\ref{fig:dmpsr} and \ref{fig:dmfrb}.
As expected, the majority of \dmdpulsar\ values are negative
with a small tail to positive values.
In contrast, the \dmdfrb\ distribution is exclusively
positive and rises sharply at 
$\mdmdfrb \approx 64 \, \mdmunits$.

We applied KDE (with Gaussian and gamma kernels, respectively) to the observed \dmdpulsar\ and \dmdfrb\ 
distributions to derive PDFs for each.
The dark, thick curves in Figures~\ref{fig:dmpsr} and
\ref{fig:dmfrb} show the results. Also overlaid on the
figures are a series of distributions derived from 1000 resampled data sets (100 shown).
Table \ref{tab:obslim} reports the final results for both models on both Galactic latitude samples. In general, we find that the uncertainty on
\dmdfrb\
values are dominated by the size of the FRB sample. However, the uncertainty on the two distributions is largely insensitive to Galactic latitude cut. 
The YMW16 model tends to have slightly smaller \dmmwism\ values for this sample, which yields larger $\max\left[\Delta\dm_\text{pulsar}\right]$ and $\min\left[\Delta\dm_\text{FRB}\right]$ estimates. However, the separation of these distributions is not sensitive to the Galactic electron density model.

\begin{table}[ht]
\captionsetup{font=footnotesize}
\footnotesize
\subfloat[]{
\hspace{-2cm}
\begin{tabularx}{\linewidth}{c|c|c|c}
 \label{tab:psrlims}
 & \multicolumn{1}{c|}{Latitude} & \multicolumn{1}{c|}{$\max\left[\mdmdpulsar\right]$} & \multicolumn{1}{c}{$\mdmmwhalo$} \\[2mm]
 \cline{1-4}
 \parbox[t]{2mm}{\multirow{2}{*}{\rotatebox[origin=c]{90}{NE2001}}}
 & $|b|>20$\,deg & $-2\pm2 \, \text{(stat)} \pm 9 \, \text{(sys)} \, \mdmunits$ & $ >-11 \, \mdmunits$\ \\[2mm]
 & $|b|>30$\,deg & $-4\pm3 \, \text{(stat)} \pm 8 \, \text{(sys)} \, \mdmunits$\ & $ >-13 \, \mdmunits$ \\[2mm]
\cline{1-4}
 \parbox[t]{2mm}{\multirow{2}{*}{\rotatebox[origin=c]{90}{YMW16}}}
 & $|b|>20$\,deg & $7\pm2 \, \text{(stat)} \pm 9 \, \text{(sys)} \, \mdmunits$ & $ >-2 \, \mdmunits$ \\[2mm]
 & $|b|>30$\,deg & $4\pm2 \, \text{(stat)} \pm 8 \, \text{(sys)} \, \mdmunits$ & $ >-5 \, \mdmunits$ \\[2mm]
\end{tabularx}
} 
\\
\subfloat[]{
\hspace{-2cm}
\begin{tabularx}{\linewidth}{c|c|c|c}
 \label{tab:frblims}
 & \multicolumn{1}{c|}{Latitude} & \multicolumn{1}{c|}{$\min\left[\mdmdfrb\right]$} & \multicolumn{1}{c}{$\mdmmwhalo$} \\[2mm]
 \cline{1-4}
 \parbox[t]{2mm}{\multirow{2}{*}{\rotatebox[origin=c]{90}{NE2001}}}
 & $|b|>20$\,deg & $54^{+40}_{-19} \, \text{(stat)} \pm 9 \, \text{(sys)} \, \mdmunits$ & $<127 \, \mdmunits$\ \\[2mm]
 & $|b|>30$\,deg & $45^{+39}_{-9}  \, \text{(stat)} \pm 7 \, \text{(sys)} \, \mdmunits$\ & $<110 \, \mdmunits$ \\[2mm]
\cline{1-4}
 \parbox[t]{2mm}{\multirow{2}{*}{\rotatebox[origin=c]{90}{YMW16}}}
 & $|b|>20$\,deg & $63^{+27}_{-21}  \, \text{(stat)} \pm 9 \, \text{(sys)} \, \mdmunits$ & $<123 \, \mdmunits$ \\[2mm]
 & $|b|>30$\,deg & $52^{+37}_{-11}  \, \text{(stat)} \pm 7 \, \text{(sys)} \, \mdmunits$ & $<113 \, \mdmunits$ \\[2mm]
\end{tabularx}
}
\caption{Constraints derived from (a) pulsar and (b) FRB observations. NE2001 and YMW16 are used to model \dmmwism\ with $|b|>20$\,deg and $|b|>30$\,deg. $\max\left[\mdmdpulsar\right]$ and $\min\left[\mdmdfrb\right]$ are calculated at $1\sigma$, and upper and lower limits for \dmmwhalo\ at 95\%\ c.l. .  Systematic errors are taken to be the difference between NE2001 and YMW16 estimates. KDE with Gaussian kernels and fixed bandwidths are used to model \dmdpulsar, and KDE with gamma kernels and varying bandwidths are used to model \dmdfrb.
}
\label{tab:obslim}
\end{table}

%%%%%%%%%%%%%%%%%%%%%%%%%%%%%%%%%%%%%%%%%%%%%%%%%%%%%%%%
\subsection{Simulated Sample}

\label{subsec:sim}
We now simulate
\dmdfrb\ to explore how the estimation of 
min[$\mdmdfrb$] is  
likely to improve as more FRB data becomes available
and to assess our choice of metric for $\rm min\left[\mdmdfrb\right]$. 
From Equation \ref{eq:frb},
\begin{equation}
\label{eq:frbsim}
    \Delta\dm_\text{FRB} = \dm_\text{MW,halo}+\dm_\text{cosmic}+\dm_\text{host} \;\; .
\end{equation}
\dmmwhalo\ has a positive minimum, whereas \dmcosmic\ and \dmhost---in principle---have minimums of zero. As such, \dmmwhalo\ provides a zero-point offset for \dmdfrb, 
i.e. $\min\left[\Delta \dm_\text{FRB}\right] > 0$. 
For the following simulation,
\dmmwhalo\ is chosen to be a delta function at 
$30$ \dmunits\, and \dmhost\ is approximated by a lognormal distribution with a mean of $\mu=40$ \dmunits\ and a standard deviation of $\sigma=0.5$. 
Other models for these quantities are explored in \S~\ref{sec:moresim}.

To generate a cosmic DM
contribution to the simulation, we must adopt a
distribution of redshifts for the FRBs.  
We choose to estimate it from 
the observed \dmfrb\ values.
Specifically, we adopt a DM--$z$ relation\footnote{Code available at \href{https://github.com/FRBs/FRB}{https://github.com/FRBs/FRB}}
to convert
the observed sample of \dmfrb\ values
to a set of redshifts. Here the observed sample set has $|b|>20$\,deg and \dmmwism\ is subtracted off with NE2001.
We then applied standard KDE with a Gaussian kernel to build a PDF of the $z$ values
from which random draws may be taken. 
The draws are fed back into the DM--$z$ relationship to obtain the average cosmic contribution to the DM, 
\begin{equation}
\langle \dm_\text{cosmic} (z) \rangle = \int \frac{\bar{n}_eds}{1+z} \;\; ,
\end{equation}
where $\bar{n}_e = f_d(z)\rho_b(z)\mu_e/\mu_mm_p$ is the average electron density, $f_d$ is the fraction of cosmic baryons in diffuse ionised gas, $\rho_b\equiv\Omega_b\rho_c$ is the cosmic baryonic mass density, and $\mu_m$ and $\mu_e$ describe properties of helium.

We allow for deviations of \dmcosmic\ from the
average value following the formalism
presented in  \cite{macquart20}.
Our treatment is simpler than theirs;  specifically,
we assume that the fractional standard deviation
of \dmacosmic\ is $\sigma_{\rm DM} = F z^{-1/2}$
with $F=0.2$. 
We may then generate a simulated \dmcosmic\
distribution based on the $z$ distribution
and random draws from a Gaussian characterized
by $\sigma_{\rm DM}=1$ and truncated at $\pm 1\sigma$.  
Throughout, we enforce $\mdmcosmic > 0$.  
The resultant \dmcosmic\ values are added to $\dm_\text{halo}$ and 
$\dm_\text{host}$ to give the simulated PDF 
of $\Delta \dm_\text{FRB}$. 

\begin{figure}[tb]
\captionsetup{font=footnotesize}
\subfloat{
	\includegraphics[width=\linewidth]{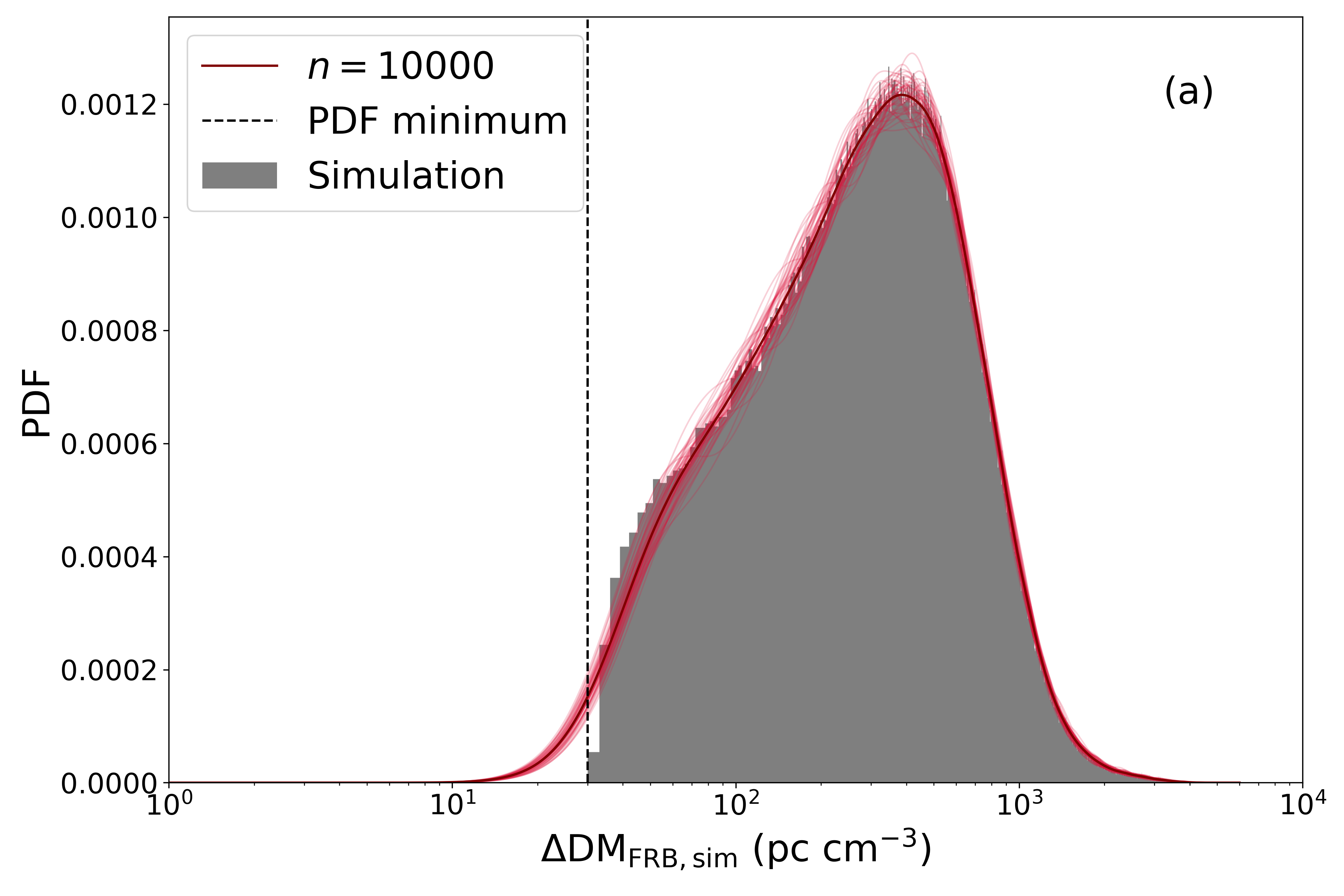}
	\label{fig:frbsim}}

\subfloat{
	\includegraphics[width=\linewidth]{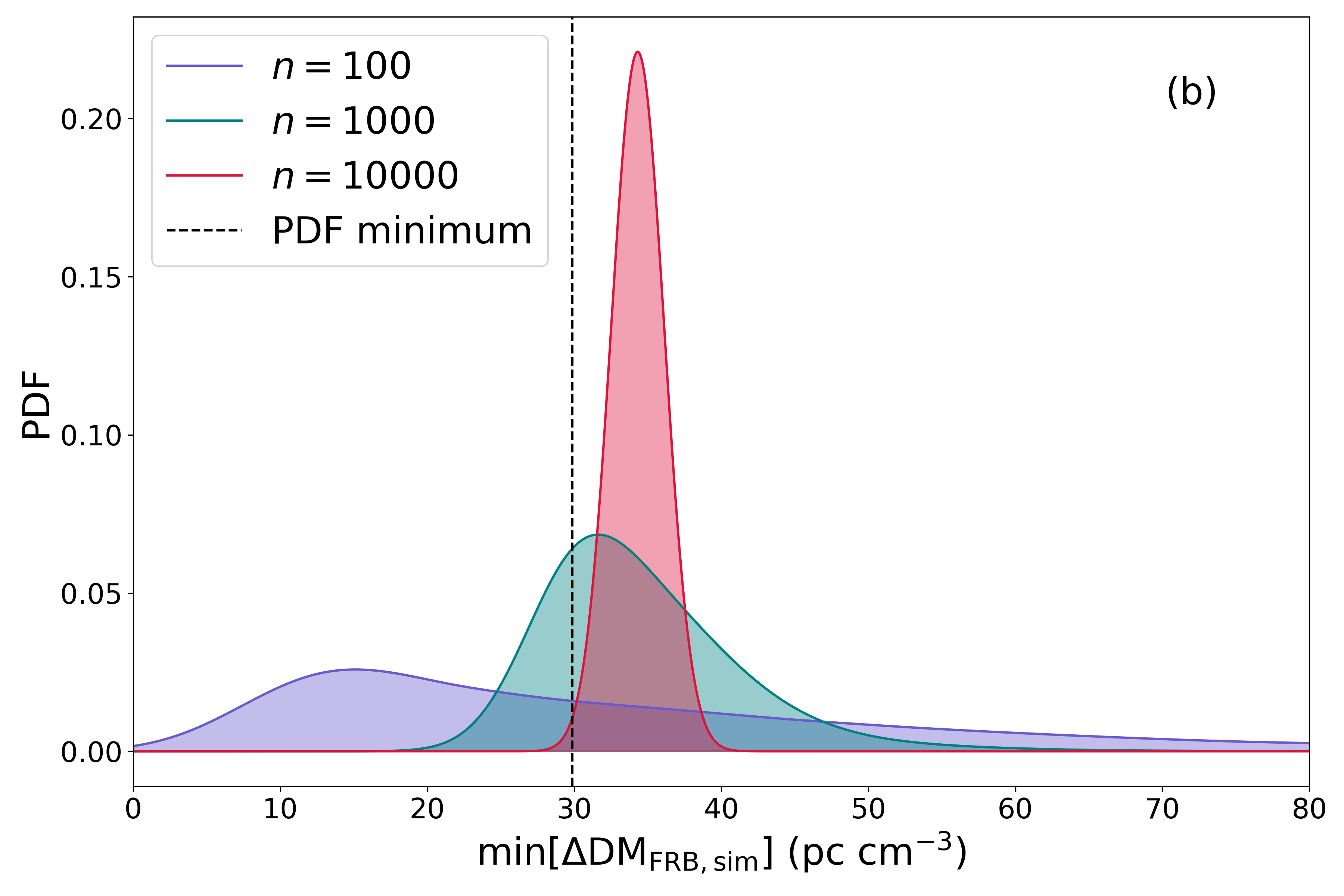}
	\label{fig:gapsim}}
\caption{(a): Distribution of $\Delta\dm_\text{FRB,sim}$ from simulated data. The KDE (gamma) estimation for $n=10,000$ is denoted by the thicker dark red line. The thinner red lines show the ensemble of KDEs from resampled data. (b): Distributions of $\min\left[\Delta\dm_\text{FRB,sim}\right]$ given by the maximum gradients of the KDE (gamma) PDFs. As the sample size increases, solutions settle with higher certainty to $\min\left[\Delta\dm_\text{FRB,sim}\right]=34\, \mdmunits$, which is $4 \, \mdmunits$ above the absolute minimum.}
\end{figure}

Figure~\ref{fig:frbsim} shows a realization of this simulated
PDF for $n=10,000$ draws.  This realization has an 
absolute minimum of $\mdmdfrb = 30 \, \mdmunits$
and rises sharply due to the host and 
\dmcosmic\ contributions. 
The dark red curve is the KDE (gamma) using the original data set and the other red curves are distributions generated with resampled data.

We explore the sensitivity of the analysis and results to
samples size $n$ as follows.  For $n=100,1000$ and 10,000,
we draw a random set of \dmdfrb\ values and model the
distributions with KDE (gamma). We then estimate a  
minimum value from the gradients of the PDFs, i.e.\ 
$\rm min\left[\mdmfrb\right]$ is the value which maximizes the slope of the KDE.
Since each $n$ PDF is complemented by 1000 PDFs resampled from the original data set, 1000 minima are available for error estimation.
The distribution of min[\dmdfrb] values are shown
in Table~\ref{tab:gapsim}.
As $n$ increases, the dispersion 
in min[\dmdfrb] decreases and the central values approach
$\approx 34 \, \mdmunits$ (Figure~\ref{fig:gapsim}). 
Adding more than 10,000 samples has no notable effect on the results.

The simulation estimates are skewed to the left for small $n$ and approach a Gaussian distribution with increased confidence as $n$ increases (Figure~\ref{fig:gapsim}). While the mean values of the distributions are similar (Table~\ref{tab:gapsim}), a sample size of $n=100$ is inadequate to place a constraint with reasonable confidence. The confidence level does however improve significantly as $n$ approaches 10,000. 

Other choices for \dmhost\ are explored to ensure the metric $\min[\Delta\dm_\text{FRB}] = \max\left[ f'(\Delta\dm_\text{FRB})\right]$ is reasonably robust to changes in the FRB simulation. Results are consistent, as detailed in \S~\ref{sec:moresim}. The smoother the leading edge of \dmdfrb\ i.e. the smoother \dmhost), the more conservative the limits become, and a very sharp edge for \dmdfrb\ i.e. a delta function for \dmhost) is described well by the metric. These cases represent extreme examples of possible host galaxy DM distributions.

\begin{table}[ht]
\captionsetup{font=footnotesize}
\centering
\footnotesize
\begin{tabularx}{\linewidth}{c|c|c}
No. FRBs & $\min\left[\Delta\dm_\text{FRB}\right]$ & $\mdmmwhalo$ \\
\cline{1-3}
100 &  $37 \pm 24 \, \text{(stat)} \, \mdmunits$ & $< 114 \, \mdmunits$  \\
1000 & $35 \pm 7  \, \text{(stat)} \, \mdmunits$ & $< 55 \, \mdmunits$ \\
10000 & $34 \pm 2  \, \text{(stat)} \, \mdmunits$ & $< 44 \, \mdmunits$ \\
\end{tabularx}
\caption{Simulation estimates for different sample sizes with  $\min\left[\Delta\dm_\text{FRB,sim}\right]=30 \, \mdmunits$. The second column gives the recovered measurements for $\min\left[\Delta\dm_\text{FRB}\right]$ at $1\sigma$ and the last column gives an upper limit for \dmmwhalo\ (95\%\ c.l.). \label{tab:gapsim}}
\end{table}

\section{Discussion}
\label{sec:discussion} 

The principle empirical result of our work is a conservative
upper limit on the DM contribution of the Milky Way halo. At $1\sigma$, $\mdmmwhalo = 63^{+27}_{-21} \, (\text{stat}) \pm 9 \, (\text{sys}) \, \mdmunits$ ($|b|>20$\,deg, YMW16). This can be converted to a conservative upper limit of $\mdmmwhalo < 123 \mdmunits$ (95\%\ c.l.).
This
includes the ISM and halo, and potentially a 
non-zero contribution from the FRB host galaxy, which is
plausibly several tens \dmunits\ (see below).
This limit also includes a non-zero contribution from 
the cosmic web (\dmcosmic). That contribution is difficult
to estimate at present but we note that the lowest redshift
FRB \citep[$z=0.03$;][]{2020Natur.577..190M} would yield an
average \dmcosmic\ of $\approx 25 \, \mdmunits$.
A more realistic, yet speculative, upper limit to 
\dmmwhalo\ may therefore be $\approx 50 \, \mdmunits$.

The results presented include two measurements of uncertainty: systematic uncertainties related to ISM models and statistical uncertainties related to the estimation techniques. Another point to consider is the effect that Galactic latitude has on results. Owing to the complexity of the electron distribution at lower Galactic latitudes, we consider cuts of $|b|>20$\,deg and $|b|>30$\,deg. Results are largely insensitive to this cut, however the loss of data at $|b|>30$\,deg (371 to 215 pulsars, and 83 to 64 FRBs), motivates a cut of $|b|>20$\,deg for our final analysis. 

Pulsar constraints are dominated by uncertainties in modeling \dmmwism. We find that, on average, \dmmwism\ values recovered from NE2001 are $\approx 10 \, \mdmunits$ lower than those from YMW16. Given the expectation that $\mdmmwhalo>0$, we use YMW16 in our analysis (see Table~\ref{tab:frblims}). This gives a final result of $\mdmmwhalo>-2 \, \mdmunits$ (95\%\ c.l.). We note that characterizing the line of sight to MW pulsars may help find HII regions that bias the \dmmwism\ estimate, allowing for improvement in the pulsar sample.

FRB constraints are predominantly limited by sample size $n$, i.e., our simulations show a significant improvement as $n$ increases. For an absolute value of $\mdmmwhalo=30 \, \mdmunits$, limits for $n=100,\, 1000$ and 10,000 are $\mdmmwhalo<114 \, \mdmunits$, $\mdmmwhalo<55 \, \mdmunits$, and $\mdmmwhalo<44 \, \mdmunits$, respectively (95\%\ c.l.). This suggests that once thousands of FRBs have been observed, the constraints will greatly improve. 

\begin{figure}[tb]
\includegraphics[width=\linewidth]{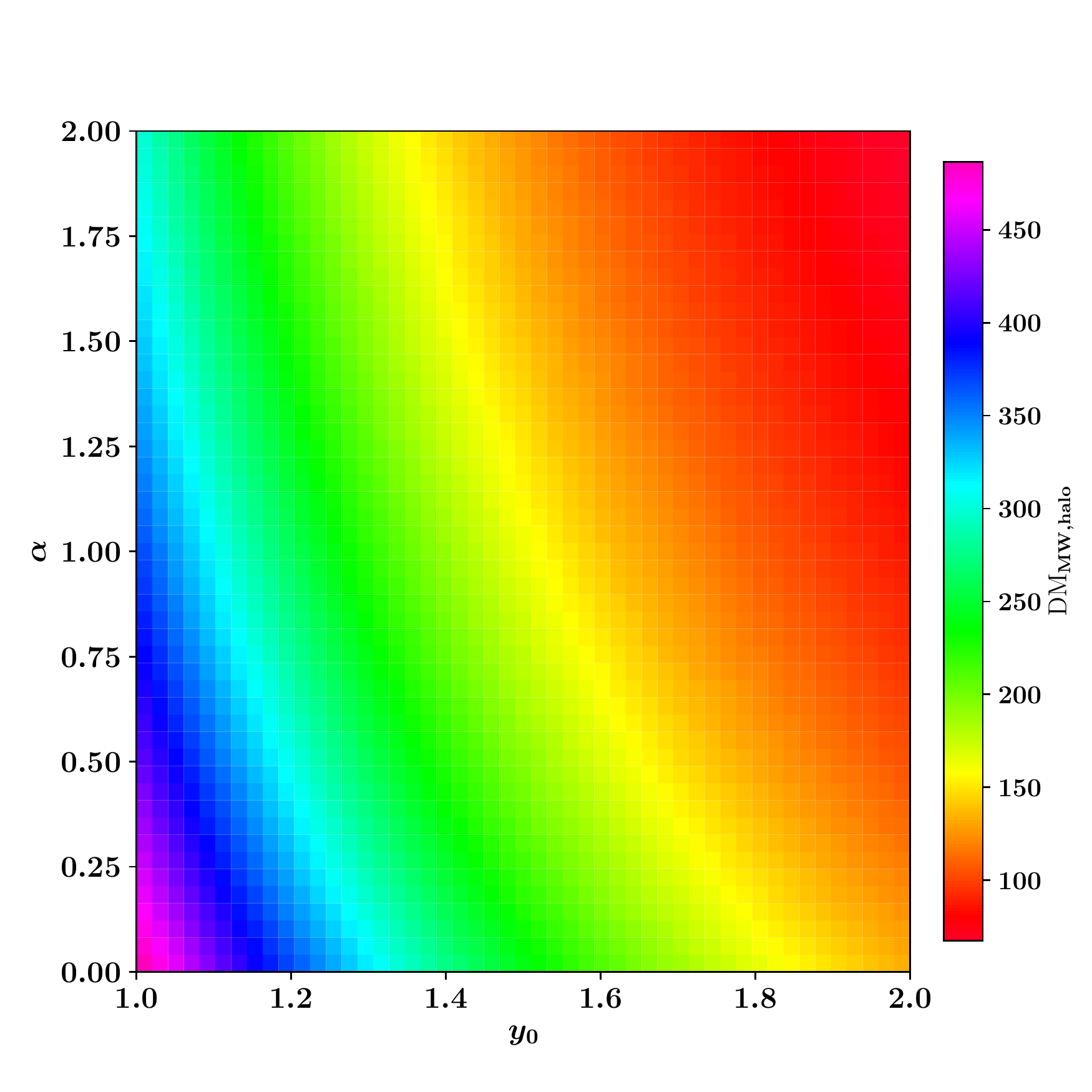}
\caption{
Predicted \dmmwhalo\ for our Galaxy as a 
function of two shape parameters that describe the
assumed baryonic density profile \citep{xyz19}.
The analysis assumes a Galactic halo with 
total baryonic mass $M_b \approx 2.4 \times 10^{11} \mmsun$
and that 75\%\ of those baryons are in 
an ionized diffuse phase of the halo.
The upper limit of $\mdmmwhalo < 123 \, \mdmunits$
rules out density profiles that more closely
resemble the NFW profile ($\alpha=0, y_0=1$).
}
\label{fig:Mb}
\end{figure}

Even the conservative limit of
$\mdmmwhalo < 123 \, \mdmunits$
offers a valuable bound to models of the Galactic halo
and the Local Group that our Galaxy resides within.
Scenarios that adopt a Galactic halo mass
$\mmhalo \approx 10^{12.2} \, \mmsun$ which has retained
all of its cosmic average of baryons
estimate $\mdmmwhalo > 50 \, \mdmunits$ 
(\citet{xyz19}, but see \citet{keating_pen2020}).
Furthermore, models which would predict the gas
traces the dark matter profile would yield 
$\mdmmwhalo > 200 \, \mdmunits$
(Figure~\ref{fig:Mb});  these are ruled out
by our FRB analysis, and also their over-estimated
X-ray emission \citep[e.g.][]{Fang:2015zya}.
Our results also place an upper bound on the average
contribution from the Local Group medium, consistent
with current estimates \citep{xyz19}.
Clearly, as the observed FRB sample increases---one expects a dramatic leap from the CHIME survey
\citep{2018ApJ...863...48C}---the resultant limits may well 
distinguish between models where the Galaxy has
retained the majority of its baryons from those 
where they have been expelled.

To illustrate the potential constraints, 
Figure~\ref{fig:Mb} shows a model-based estimate
for \dmmwhalo\ for a dark matter halo with
mass $\mmhalo = 10^{12.2} \mmsun$, 
baryonic mass $M_b = \Omega_b/\Omega_m \mmhalo 
\approx 2.4 \times 10^{11} \mmsun$
and that $75\%$ of those baryons are in a diffuse,
ionized halo.
The density profile is assumed to follow a modified
Navarro-Frenk-White (NFW) profile parameterized by $y_0$ and $\alpha$
\citep[see][]{mp2017,xyz19}. 
The upper limit to \dmmwhalo\ estimated from our analysis
prefers larger $\alpha, y_0$
with a strict NFW profile ($\alpha=0, y_0=1)$
ruled out at high confidence unless
$M_b \ll \Omega_b/\Omega_m \mmhalo$.
Larger $\alpha, y_0$ are inferred for our Galaxy
and external ones from absorption-line
analyses \citep[e.g.][]{ovi,mp2017}.

We emphasize that ongoing FRB projects will offer complementary
constraints on the magnitude and distribution of contributions
from the host and the cosmic web to the upper limit on \dmmwhalo.
In particular, well-localized FRBs reveal the host galaxy population
and the redshift distribution of FRB events.  From follow-up
observations of the hosts, one may estimate the DM contribution
from the host galaxy ISM through measurements of the Balmer line
emission \citep[e.g.][]{Tendulkar:2017vuq,chittidi2020}.
The two systems analyzed thus far yield 
${\rm DM}_{\rm host,ISM} \approx 50$\textendash$200 \, \mdmunits$.
There are other FRBs \citep[e.g.\ FRB~180924; ][]{2019Sci...365..565B}
where the Balmer emission is low or even negligible at the FRB
location and we infer ${\rm DM}_{\rm host,ISM} < 50 \, \mdmunits$.
Within the next year, we expect to have a sample of $\sim 20$ hosts
to derive the distribution.

One may additionally translate the estimated stellar mass of the
host galaxy into a model-based estimate for the DM contribution
from the halo gas of the host \citep{2019Sci...365..565B,xyz19}.
Current estimates range from $\approx 50 \, \mdmunits$ for the most
massive hosts \citep{2019Sci...365..565B}
to $<20 \, \mdmunits$ for FRB~181112 \citep{xyz19}.
From the redshift distribution of the localized FRBs, one may 
estimate the minimum typical contribution of \dmcosmic\ to 
the \dmmwhalo\ limit.  This bears an important caveat that the
selection biases of the localized sample will not match those
of the larger ensemble (e.g.\ due to differences in the radio
frequencies and/or flux limit).
One will need to account for these differences.
Alternatively, one may focus on the analysis of the a localized
sample alone once it grows to a sufficient sample size.
 
Last, we emphasize that other, future observations will also offer
constraints on \dmmwhalo\ independent of FRB analyses.
We anticipate high-precision X-ray absorption-line spectroscopy
of the Galactic halo from the upcoming Japanese XRISM mission.
With a spectral resolution that will greatly exceed current 
X-ray satellites, the data will yield much more reliable estimates
of O$^{+5}$ and O$^{+6}$ column densities across the sky.
At the least, these yield conservative lower limits to 
\dmmwhalo.  Another promising yet still unrealized opportunity
is to discover pulsars in Andromeda or any other Local Group 
galaxy.  These would offer a strict upper bound on \dmmwhalo\ 
or even a well-informed value along that sightline.

%%%%%%%%%%%%%%%%%%%%%%%%%%%%%%%%%%%%%%%%%%%%%%%%%%%%%%%%%%%
\section{Concluding Remarks}
\label{sec:conclusion}

We have demonstrated how density estimation techniques can be used to probe the DM---i.e. the line-of-sight electron column density---of the MW Galactic halo. For the corrected \dmpulsar\ and \dmfrb\ distributions, we recover $\max\left[\dm_{\rm pulsar}\right] \approx 7\pm2 \, \text{(stat)} \pm 9 \, \text{(sys)} \, \mdmunits$ and $\min \left[\dm_{\rm FRB}\right] \approx 63^{+27}_{-21}  \, \text{(stat)} \pm 9 \, \text{(sys)} \, \mdmunits$ ($1\sigma$ uncertainty). Conservative upper and lower limits on the Galactic halo dispersion measure are also derived:
$\mdmmwhalo >-2 \, \mdmunits$ and  $\mdmmwhalo <123 \, \mdmunits$ (95\%\ c.l.).
Here the lower bound given by pulsars reflects only a fraction of the MW halo DM, and the upper bound given by FRBs includes a nominal contribution from the FRB host galaxy and IGM. In the latter case, the localization of FRBs at very low distances and/or on the outskirts of galaxies would establish that the minimum DM would be more representative of the MW halo. Scenarios consistent with this include the collapse of compact objects \citep[e.g.][]{Falcke:2013} that have been expelled from a host galaxy, as well as more exotic theories such as tiny electromagnetic explosions \citep[which may occur in dark matter halos;][]{Thompson:2017inw,Thompson:2017wbo} and cosmic strings \citep[e.g.][]{Vachaspati:2008su,Yu:2014gea,2015AASP....5...43Z,Brandenberger:2017}.

We do not consider how \dmmwhalo\ may vary as a function of Galactic latitude. It may be possible with a sample of a couple thousand FRBs per region of sky, but is left to future work.

Our current estimates cannot yet discern whether the Milky Way has retained its cosmic average of baryons ($\mdmmwhalo > 50 \, \mdmunits$),
however in the near future, as more FRBs are reported, results may offer a valuable complement to other analyses. In the least, the methodology provides a reasonable---albeit conservative---estimate of \dmmwhalo\ and a minimum contribution from \dmhost. This may discern the viability of Galactic halo models and aid in the search for missing baryons.

%%%%%%%%%

\acknowledgments
We would like to thank the anonymous referee for their insightful, thorough and valuable input.
EP and JXP, as members of the Fast and Fortunate for FRB
Follow-up team \href{http://www.ucolick.org/f-4}{F$^4$}, acknowledge support from NSF grant AST-1911140.
CJL acknowledges support under NSF grant 2022546. This work was initiated as a project for the Kavli Summer Program in Astrophysics held at the University of California, Santa Cruz in 2019. The program was funded by the Kavli Foundation, The National Science Foundation, UC Santa Cruz, and the Simons Foundation. We thank them for their generous support. EP is supported by a L'Or\'{e}al-UNESCO For Women in Science Young Talents Fellowship, by a PhD fellowship from the South African National Institute for Theoretical Physics (NITheP), and by a top-up bursary from the South African Research Chairs Initiative of the Department of Science and Technology (SARChI) and the National Research Foundation (NRF) of South Africa. Any opinion, finding and conclusion or recommendation expressed in this material is that of the authors and the NRF does not accept any liability in this regard.

\bibliographystyle{aasjournal}
\bibliography{dmhalo}

\appendix

\section{Minimum of FRB DM Distribution}
\label{sec:moresim}
We postulate that the minimum of the \dmdfrb\ distribution can be approximated by $\min[\Delta\dm_\text{FRB}] = \max\left[ f'(\Delta\dm_\text{FRB})\right]$. This metric is based on the prior that the underlying distribution has a sharp leading edge and is motivated by simulations. To bear weight, the metric must hold for a wide range of reasonable  $\Delta\dm_\text{FRB}$ distributions.

\begin{figure}[h]
    \centering
    \subfloat{{\includegraphics[width=8cm]{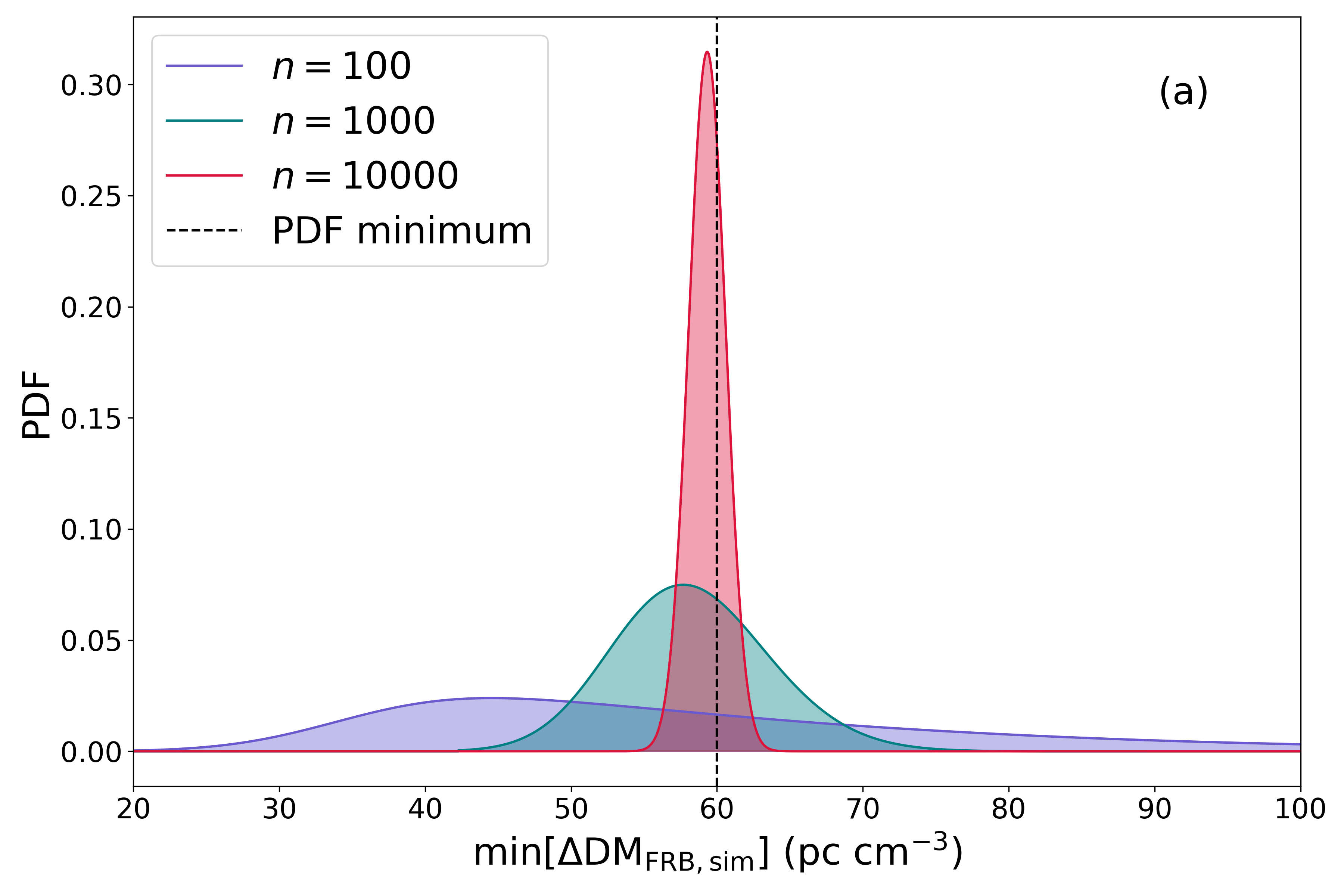} \label{fig:delta}}}%
    \qquad
    \subfloat{{\includegraphics[width=8cm]{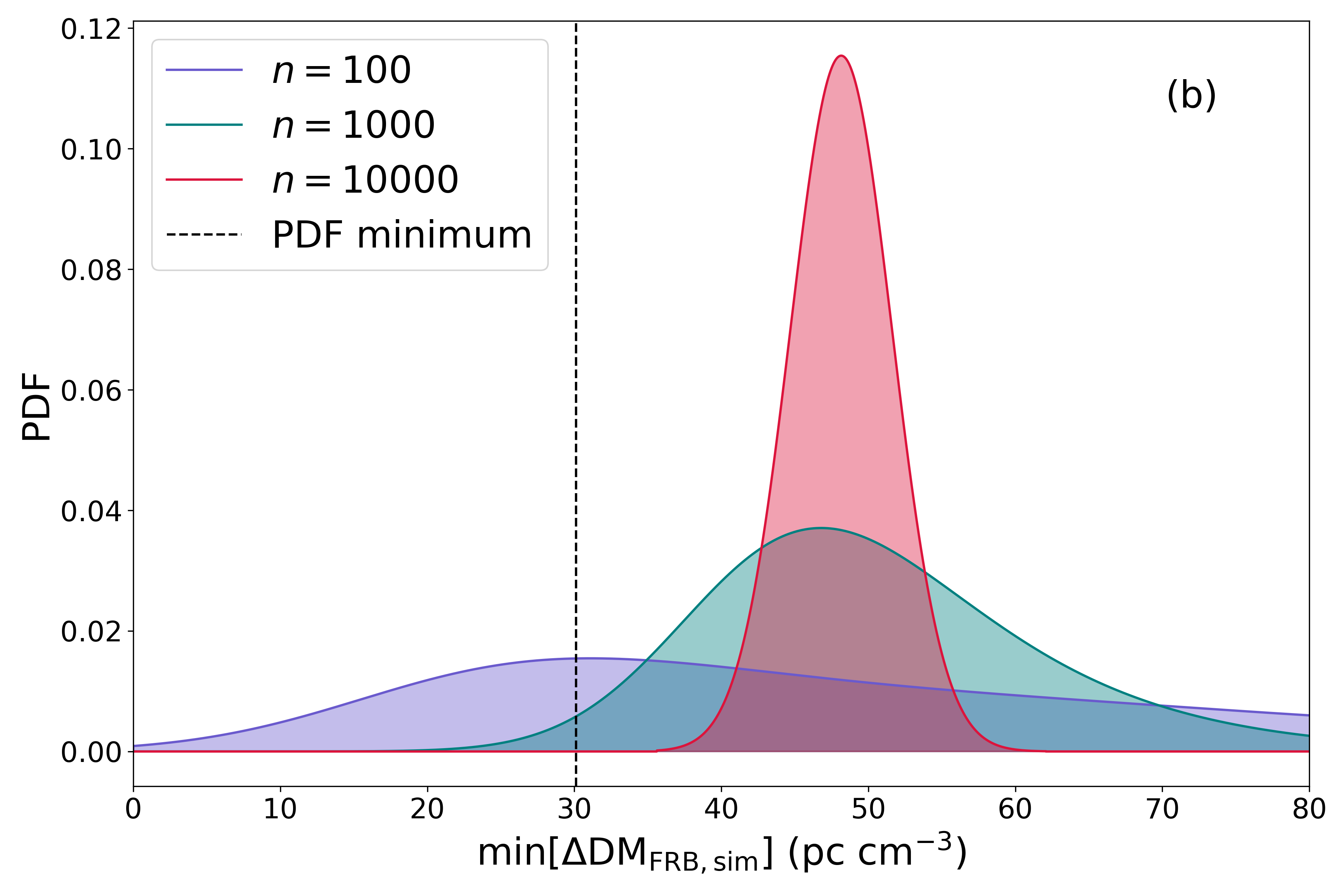} }\label{fig:gauss}}%
    \caption{(a) $\min\left[\mdmdfrb \right]$ with \dmhost\ a delta function at 30 \dmunits. The absolute minimum is 60 \dmunits. (b) $\min\left[\mdmdfrb \right]$ for a Gaussian \dmhost\ with $\mu=60 \, \mdmunits$ and $\sigma=15$. The absolute minimum is 30 \dmunits.}
    \label{fig:deltagauss}%
\end{figure}

The MW can be given by a delta function (as its DM is thought to vary by 10 \dmunits) and the cosmic DM distribution can be modelled theoretically. The distribution of host galaxy DMs, however, is unknown. In the main analysis we consider a lognormal distribution with $\mu=40$ \dmunits\ and a standard deviation of $\sigma=0.5$. Here we consider two extreme variations: a delta function at $30 \mdmunits$ and a broad Gaussian distribution with $\mu=60 \, \mdmunits$ and $\sigma=0.5$. The former distribution makes the edge of \dmdfrb\ sharper and the latter makes it smoother. The metric is a reasonable approximation for the combined $\mdmmwhalo+\mdmhost$ contribution when each distribution is sharp (Figure~\ref{fig:delta}). When \dmhost\ has a smooth edge, the estimates are more conservative (Figure~\ref{fig:gauss}). Thus, provided the leading edge of \dmdfrb\ is sufficiently sharp, the metric for determining the distribution minimum can be considered reasonably robust.

Looking at Figure~\ref{fig:deltagauss}, a sample size of $n=1000$ appears sufficient to provide an estimate consistent with that of $n=10,000$. For $n=100$, distributions are wide and skewed to the left, providing results that are clearly premature.

\section{Density Estimation Using Field Theory}
\label{sec:deft}
Density estimation using field theory \citep[DEFT;][]{PhysRevE.90.011301,PhysRevE.92.032107,PhysRevLett.121.160605} is a newly developed technique specifically developed for the small data regime. It takes a Bayesian field theory approach to density estimation in small data sets using a Laplace approximation of the Bayesian posterior (also see \citet{riihimaki2014}). An advantage of DEFT over standard density estimation methods is that the method does not require the manual identification of critical parameters nor does it require the specification of boundary conditions. The DEFT simulations in this paper use the Python package \code{SUFTware} (Statistics Using Field Theory) by \citet{PhysRevLett.121.160605}.

Consider $n$ data points ($x_1, x_2,...,x_n$) drawn from a known probability distribution $Q_{\text{true}}(x)$ with $x$ intervals of length $L$. We wish to find the best estimate $Q^*(x)$ of this distribution and the  accompanying ensemble of other plausible estimates. Each distribution  $Q(x)$ is parameterized by a real field $\phi(x)$, ensuring that $Q(x)$ is positive and normalized:
\begin{equation}\label{key}
Q(x) = \frac{e^{-\phi(x)}}{\int dx'e^{-\phi(x')}} \;\; .
\end{equation}
Using scalar field theory, a prior $p(\phi | \ell)$ is formulated that favours smooth probability densities. Specifically, \citet{PhysRevE.92.032107} consider priors of the form
\begin{equation}
p(\phi | \ell) = \frac{e^{-S_\ell^0\left[\phi\right]}}{Z_\ell^0}  \;\; ,
\end{equation}
with action

\begin{equation}
S_\ell^0\left[\phi\right] = \int \frac{dx}{L}\frac{\ell^{2\alpha}}{2}\left(\partial^\alpha\phi\right)^2  \;\; ,
\end{equation}
and partition function

\begin{equation}
Z_\ell^0 = \int \mathcal{D}\phi e^{-S_\ell^0[\phi]} \;\; .
\end{equation}
Here, $\ell$ gives the length scale below which $\phi$ fluctuations are strongly damped and $\alpha>0$ is an integer in the range [1,...,4] that determines the smoothness. The resultant posterior is given by

\begin{equation}
p(\phi | \text{data}, \ell) = \frac{e^{-S_\ell\left[\phi\right]}}{Z_\ell}  \;\; ,
\end{equation}
with nonlinear action
\begin{equation}
S_\ell\left[\phi\right] = \int \frac{dx}{L}\left\{ \frac{\ell^2\alpha}{2} \left(\partial^\alpha\phi\right)^2 +nLR\phi +ne^{-\phi} \right\} \;\; ,
\end{equation}
and partition function
\begin{equation}
Z_\ell = \int \mathcal{D}\phi e^{-S_\ell[\phi]} \;\; .
\end{equation}
$R(x)=\frac{1}{n}\sum^n_{i=1}\partial (x-x_i)$ is a histogram that summarizes the data.

Maximum \textit{a posteriori} (MAP) density estimation approximates the posterior $p(\phi | \text{data},\ell)$ as a $\delta$ function given by the mode of the posterior, at which the action $S_\ell\left[\phi\right]$ is then minimized. It has been shown that even without imposing boundary conditions on $\phi$, $S_\ell\left[\phi\right]$ has a unique minimum \citep{PhysRevE.92.032107}. The optimal length scale $\ell^*$ is identified by maximizing the Bayesian evidence $p(\text{data}| \ell)$.

The uncertainty in the DEFT estimate $Q^*$ is determined by  sampling the Bayesian posterior,
\begin{equation}
p(Q|\text{data}) = \int dlp(\ell|\text{data})p(Q|\text{data},\ell) \;\; ,
\end{equation}
by first drawing $\ell$ from $p(\ell|\text{data})$ and then drawing $Q$ from $p(Q|\text{data},\ell)$. Laplace approximation is used to estimate $p(Q|\text{data},\ell)$ by constructing a Gaussian centered at its MAP value. This gives the Laplace posterior,
\begin{equation}
p_{\text{Lap}}(Q|\text{data}) = \int 
dlp(\ell|\text{data})p_{\text{Lap}}(Q|\text{data},\ell) \;\; ,
\end{equation}
from which an ensemble of distributions $Q$ can be sampled. Some of the $Q$s, however, are clearly not representative of the underlying distribution. Importance resampling is thus used to remove unfavorable distributions, where each $\phi$ is given a weight,
\begin{equation}
w_\ell[\phi]=\exp \left(S^{\text{Lap}}_\ell[\phi] -S_\ell[\phi] \right) \;\; ,
\end{equation}
proportional to its probability of being drawn \citep{PhysRevLett.121.160605}. DEFT uses importance resampling with replacement, however for this work we invoke importance resampling without replacement.

When a posterior turns out to be a poor approximation of the target distribution, a few of the sampled distributions are given very large weights and the majority are given small weights \citep{Gelman:1995,Skare:2003}. When resampling with replacement, the heavily weighted distributions become significantly over represented. In our case, $\sim$ 60--70\% of the sampled distributions were duplicates, which lead to notable bias when calculating the upper and lower bounds of \dmmwhalo. As such, we use a set of the most probable distributions, with limited replications. Specifically, we select 500 out of 1000 distributions via importance sampling without replacement. This lowered the duplication rate to $\sim10\%$.

\begin{figure}[tb]
\centering
\subfloat{
	\includegraphics[width=0.5\linewidth]{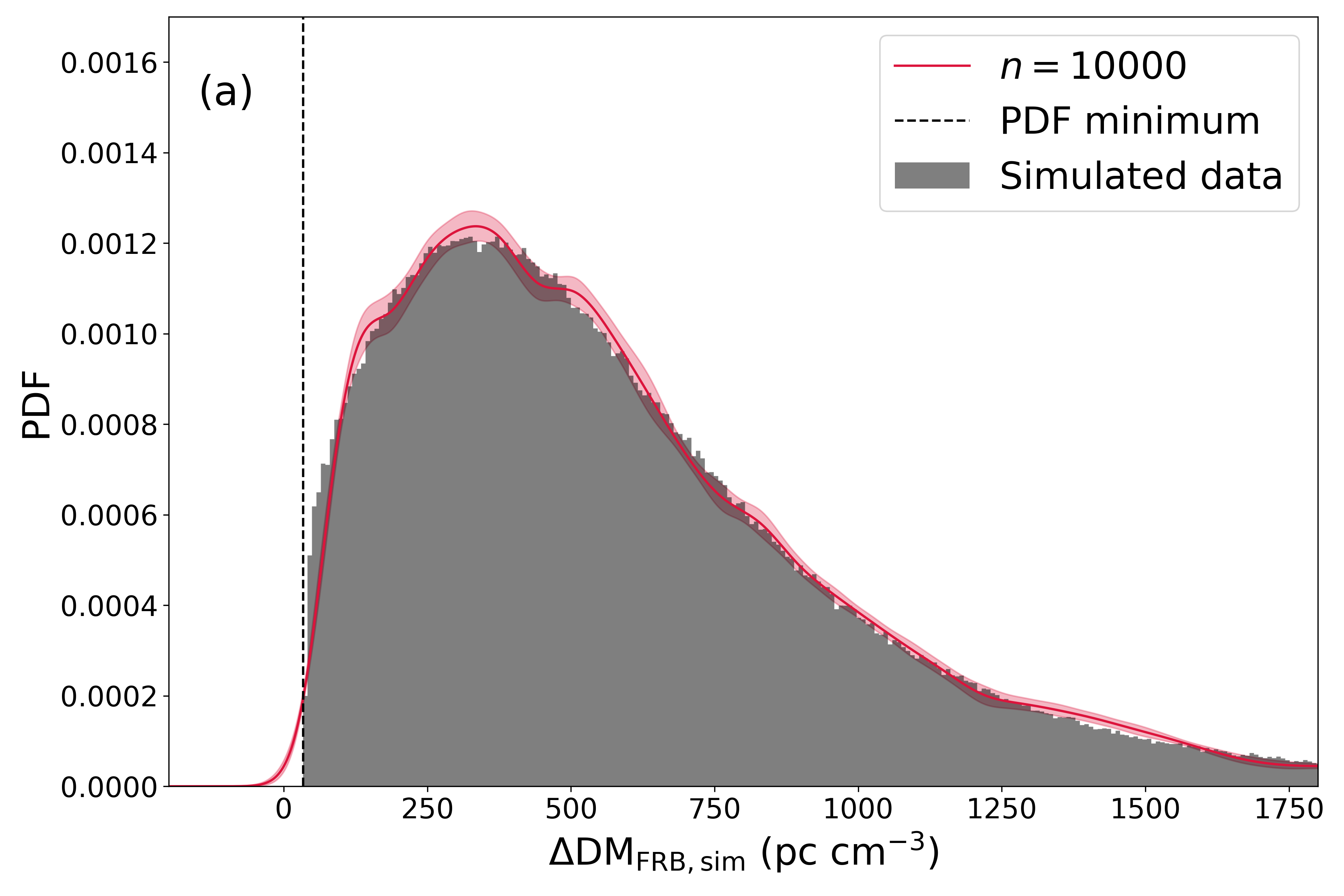}
	\label{fig:deft}}
	
\caption{
Distributions of \dmdfrb\ for 10,000 samples, restricted to $|b| > 20$\,deg and using
NE2001 for modeling \dmmwism.  
Overlaid on the data are PDFs derived with DEFT. The thick line denotes the DEFT Bayesiean posterior and shaded line denotes standard deviation of the set of PDFs derived by sampling the Bayesiean posterior.
}
\end{figure}

We approximate the FRB distribution described in \S \ref{subsec:sim} using DEFT for $n=100$, $n=1000$ and $n=10,000$. Even for large $n$ DEFT is unable to adequately describe the sharp edge of the simulated distribution. In Figure \ref{fig:deft}, the PDF tail extends below zero, violating the physical condition that \dmdfrb$>0$. Further, the PDF cuts straight through the front of the simulated distribution and so bypasses the structure of the distribution's edge. 

\section{Generalized Extreme Value}
\label{sec:gev}

A standard statistical technique for estimating the maximum 
values of an ensemble to fit it with a Generalized
Extreme Value (GEV) PDF \citep[e.g.][]{cole}.
This technique, however, is most applicable for assessing
the upper limit of a distribution with a long tail.  
For \dmdfrb,  this holds for the largest values but the
lowest values rise sharply as one may expect 
from the MW and host contributions.

Nevertheless, we attempted to estimate the minimum of
\dmdfrb\ following the standard practice of assessing
the maximum of the negative of the distribution
\citep{cole}.  The results reported a minimum value
at effectively infinite confidence at the lowest 
\dmdfrb\ in the distribution and we found the results
were unstable to random sampling.   

\end{document}